\documentclass[aps,twocolumn,showpacs,showkeys,amsmath,amssymb,pra]{revtex4} 

\usepackage{graphicx}   
\usepackage{dcolumn}    
\usepackage{bm}         
\usepackage{textcomp}	
\usepackage{natbib}

\newcommand{\rd}{{\rm d}} 
 
\newcommand{\ri}{{\rm i}} 
\newcommand{\rE}{{\rm E}}
\newcommand{\rH}{{\rm H}}
\newcommand{\kB}{k_{\rm B}} 
\newcommand{\Tp}{T_{\rm P}}
\newcommand{\Ts}{T_{\rm S}}
\newcommand{\tr}{{\rm tr}}
\newcommand{\Ip}{{\rm Im}}
\newcommand{\oth}{\omega_{\rm th}}

\begin{document}
   
\title{Spheroidal nanoparticles as thermal near-field sensors}

\author{Svend-Age Biehs\footnote{Present address: Laboratoire Charles Fabry,  
	Institut d'Optique, CNRS, Universit\'{e} Paris-Sud, 
	Campus Polytechnique, RD128, 91127 Palaiseau cedex, France}}
\author{Oliver Huth}
\author{Felix R\"uting}
\author{Martin Holthaus}
\affiliation{Institut f\"ur Physik, Carl von Ossietzky Universit\"at, 
	D-26111 Oldenburg, Germany}
\date{April 20, 2010}

\begin{abstract}
We suggest to exploit the shape-dependence of the near-field heat transfer
for nanoscale thermal imaging. By utilizing strongly prolate or oblate 
nanoparticles as sensors one can assess individual components of the
correlation tensors characterizing the thermal near field close to a
nanostructured surface, and thus obtain directional information beyond the 
local density of states. Our theoretical considerations are backed by 
idealized numerical model calculations.        
\end{abstract}




\maketitle
 
\section{Introduction} 
 
While profound theoretical understanding of thermally generated electromagnetic
fields close to material surfaces has already been obtained some time 
ago~\cite{RytovEtAl89,PolderVanHove71,JoulainEtAl05,VolokitinPersson07}, 
accurate experimental studies of such fluctuating thermal near fields at 
distances on the order of or even significantly shorter than the thermal 
wave length from a sample's surface have only recently become possible. 
Hu {\em et al.\/} have measured the radiative heat flux between two glass 
plates spaced by a micron-sized gap~\cite{HuEtAl08}, and reported near-field 
heat transfer exceeding the far-field limit set by Planck's law of blackbody 
radiation. Chen and co-workers have studied the near-field heat flux between 
a sphere and a flat substrate~\cite{NarayanaswamyEtAl08,ShenEtAl09}, 
focusing on the role of surface phonon polaritons, and observed heat transfer 
coefficients even three orders of magnitude larger than the blackbody 
radiation limit. Rousseau {\em et al.\/} have made precise measurements of the 
radiative heat transfer between spheres with diameters of 22 and 40 \textmu m 
at distances from 30~nm to 2.5~\textmu m from a plate~\cite{RousseauEtAl09}, 
and obtained impressive agreement with predictions based on fluctuational 
electrodynamics~\cite{RytovEtAl89,PolderVanHove71}. This finding 
is conceptually important since it indicates that fluctuational 
electrodynamics, which is a macroscopic theory, can be relied on 
at least up to the smallest distances accessible in this experiment, 
without having to account for effects caused by a nonlocal optical 
response~\cite{HenkelJoulain06,ChapuisEtAl08B}. Moreover, a device termed 
Near-Field Scanning Thermal Microscope (NSThM) has been designed by Kittel 
{\em et al.\/}~\cite{MuellerHirschEtAl99,KittelEtAl05,WischnathEtAl08}
This instrument allows one to record the near-field heat flux at probe-sample 
distances even down to a few nanometers, but still further efforts are 
required to achieve accurate calibration.  A complementary set-up dubbed 
Thermal Radiation Scanning Tunneling Microscope (TRSTM) has enabled 
De Wilde {\em et al.\/} to take images of thermally excited surface plasmons, 
and to demonstrate spatial coherence effects in near-field thermal 
emission~\cite{DeWildeEtAl06}. This notable growth of experimental research 
on thermal near-field phenomena is triggered, on the one hand, by a wide 
variety of possible technological applications, ranging from the design 
of nanoscale heaters for use in heat-assisted magnetic recording or 
heat-assisted lithography~\cite{RousseauEtAl09} to near-field 
thermophotovoltaics~\cite{DiMatteoEtAl01,NarayanaswamyChen03,LarocheEtAl06,
FrancoeurEtAl08,BasuEtAl09}. On the other hand, there is a need for 
further insight into basic small-scale phenomena. For example, at very 
short distances nonlocal effects ultimately must come into
play~\cite{HenkelJoulain06,ChapuisEtAl08B}, and the use of macroscopic 
fluctuational electrodynamics as a theoretical framework might no longer be 
sufficient. Thus, thermally induced near-field effects constitute an emerging 
subject which requires further experimental and theoretical investigations; 
of particular importance is the identification of experimentally observable 
quantities against which theoretical models can be tested.  

It has already been demonstrated experimentally that the NSThM opens up 
the possibility of thermal imaging of surface structures with nanoscale 
resolution~\cite{KittelEtAl08}, since the surface topography leaves its 
imprint in the local density of states (LDOS). In this paper we take a further 
step in this direction: We argue that one can obtain information even beyond 
the mere topography if the NSThM sensor is appropriately shaped. This is 
due to the fact that in principle not only the LDOS, but even the individual
diagonal components of the fluctuating fields' correlation tensors are 
measurable; these components are singled out when employing strongly prolate 
or oblate spheroidal particles as sensors. Therefore, when scanning a surface 
at nanometer distances with such a sensor, the NSThM-signal does not give 
a one-to-one image of the surface profile, but rather provides information on
how that profile affects the directional properties of the fluctuating thermal 
near field. We first sketch the theoretical idea underlying this proposal in 
Sec.~\ref{S_2}, and then present in Sec.~\ref{S_3} some numerical model 
calculations which illustrate its principal feasibility. These calculations 
necessarily are strongly idealized, and do not refer to an already existing 
experimental set-up; nonetheless, they indicate a possible direction for 
future developments.

\section{The basic principle}
\label{S_2}       
 
Consider a substrate with temperature $\Ts$. The electric field 
$\mathbf{E}(\mathbf{r},t)$ close to its surface, generated by thermal 
fluctuations inside the sample, then is characterized in terms of its 
correlation tensor $W^\rE_{jk}(\mathbf{r},\mathbf{r}',\omega)$, which is 
determined by an ensemble average        
\begin{equation}
	\langle E_j(\mathbf{r},\omega) E^*_k(\mathbf{r}',\omega') \rangle
	= W^\rE_{jk}(\mathbf{r},\mathbf{r}',\omega) \, 
	\delta(\omega - \omega') \; ,
\end{equation}
where $E_j(\mathbf{r},\omega)$ are the field's Fourier components. The
magnetic correlation tensor $W^\rH_{jk}(\mathbf{r},\mathbf{r}',\omega)$ 
is defined in an analogous manner. According to fluctuational
electrodynamics~\cite{RytovEtAl89,PolderVanHove71}, these correlation 
tensors are proportional to the Bose-Einstein function evaluated at the
temperature of the sample,
\begin{equation}
	\Theta(\omega,\Ts) = \frac{\hbar\omega}{2}
	+ \frac{\hbar\omega}{\exp[\hbar\omega/(\kB\Ts)] - 1} \; .
\end{equation}
Now assume that a nanoparticle with electric polarizability tensor
$\alpha^\rE(\omega)$ is brought into this fluctuating near field at a 
position $\mathbf{r}$, such that its distance from the surface still remains 
appreciably larger than its characteristic linear size, while that size 
should be smaller than the thermal wavelength at $\Ts$. Then the field 
introduces a dipole moment $\mathbf{p}(t)$ in the particle which in its 
turn interacts with the field, such that the power    
\begin{eqnarray}
\label{eq:PDE}
	& & \langle \dot{\mathbf{p}}(t) \cdot \mathbf{E}(t) \rangle
\\
	& = & \int_{-\infty}^{+\infty} \! \rd \omega \, 
	(-\ri \omega) \frac{\varepsilon_0}{(2\pi)^2} 
	\tr \Big( 
	\alpha^\rE(\omega) \cdot W^\rE(\mathbf{r},\mathbf{r},\omega)
	\Big) 
\nonumber		
\end{eqnarray} 
is dissipated in the particle, resulting in its heating. The right-hand side 
of this equation still is written in a manner which shows that it does not 
depend on the choice of the particular coordinate system used for its 
evaluation, as the trace over the product of the polarization tensor 
and the correlation tensor manifestly remains invariant under coordinate 
transformations. When adopting a coordinate system in which the tensor 
$\alpha^\rE(\omega)$ is diagonal, with diagonal elements   
$\alpha^\rE_k(\omega)$, this expression~(\ref{eq:PDE}) takes the more 
familiar form~\cite{Dorofeyev98,Pendry99,DedkovKyasov07,ChapuisEtAl08}
\begin{eqnarray}
	& & \langle \dot{\mathbf{p}}(t) \cdot \mathbf{E}(t) \rangle 
\\	
	& = & 2 \sum_{k=1}^3 \int_0^{+\infty} \! \rd \omega \, \omega \,  
	\Ip\big(\alpha^\rE_k(\omega)\big) \frac{\varepsilon_0}{(2\pi)^2} 
	W^\rE_{kk}(\mathbf{r},\mathbf{r},\omega) \; .
\nonumber
\end{eqnarray}
It is essential to keep in mind that here the diagonal components 
$W^\rE_{kk}(\mathbf{r},\mathbf{r},\omega)$ of the sample's field correlation  
are given in the principal-axis system of the particle's polarizability, not 
in a system attached to the sample's geometry or orientation. If all elements 
$\alpha^\rE_k(\omega)$ are equal, as in the case of a nanosphere, the 
integrand simply is proportional to the trace of the correlation tensor, 
{\em i.e.\/}, to the spectral energy density   	 
\begin{equation}
	\langle u^\rE(\mathbf{r},\omega) \rangle
	= \frac{\varepsilon_0}{(2\pi)^2} 
	\sum_{k=1}^3 W^\rE_{kk}(\mathbf{r},\mathbf{r},\omega) \; ,
\label{eq:EDE}
\end{equation}
which is the case considered usually~\cite{Dorofeyev98,Pendry99,DedkovKyasov07,
ChapuisEtAl08}. If, however, the elements $\alpha^\rE_k(\omega)$ differ 
significantly from each other, the dissipated power can be dominated by 
individual components $W^\rE_{kk}(\mathbf{r},\mathbf{r},\omega)$. This is 
the principle we are going to explore.

\begin{figure}[tb]
\includegraphics[scale=0.25,angle=-90]{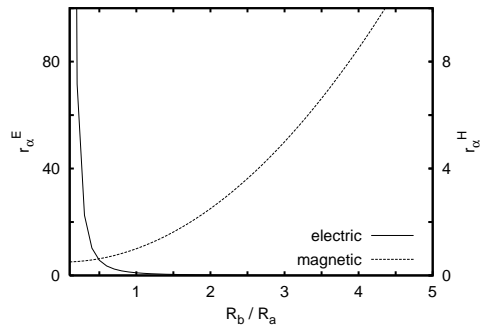}
\caption{Ratios $r_{\alpha}^{\rE/\rH} \equiv 
	\Ip(\alpha_{\rm a}^{\rE/\rH}) / \Ip(\alpha_{\rm b}^{\rE/\rH})$ 
	for a spheroidal gold nanoparticle at $\oth = 10^{14}$~s$^{-1}$.
	The particle's volume is kept constant when the aspect ratio
	$R_{\rm b}/R_{\rm a}$ is varied.} 
\label{fig:F_1}		
\end{figure}

For obtaining a convenient expression for the net power flow  
$P_{\rm S \leftrightarrow P}$ between the nanoparticle and the substrate,
we factorize the correlation tensor in the form
\begin{equation}
	\frac{\varepsilon_0}{(2\pi)^2}	
	W^\rE_{kk}(\mathbf{r},\mathbf{r},\omega)
	= \Theta(\omega,\Ts) \, D^\rE_{kk}(\mathbf{r},\omega) \; ,	
\label{eq:DKK}
\end{equation}
so that the quantities $D^\rE_{kk}(\mathbf{r},\omega)$ carry the same dimension
(sm$^{-3}$) as the local density of states (LDOS)~\cite{JoulainEtAl03}. 
Then two additional mechanisms have to be accounted for: On the one hand, 
as strongly emphasized by Dedkov and Kyasov~\cite{DedkovKyasov07} and by 
Chapuis {\em et al.\/}~\cite{ChapuisEtAl08}, the magnetic part 
$\mathbf{H}(\mathbf{r},t)$ of the fluctuating near field gives rise to 
the induction of an effective magnetic dipole $\mathbf{m}_{\rm eff}(t)$
in the particle, and hence to a further contribution 
$\langle \dot{\mathbf{m}}_{\rm eff}(t) \cdot \mathbf{H}(t) \rangle$
to the dissipated energy. On the other hand, there also is a flow of energy 
from the particle, with temperature $\Tp$, to the sample. In total, this 
results in the formula
\begin{eqnarray}
\label{eq:PSP}
	P_{\rm S \leftrightarrow P} & = &
	2 \sum_{k=1}^3 \int_0^{+\infty} \! \rd \omega \, \omega
	\Big( \Theta(\omega,\Ts) - \Theta(\omega,\Tp) \Big)
\\ & \times & 	
	\Big( 
	\Ip\big(\alpha^\rE_k(\omega)\big) D^\rE_{kk}(\mathbf{r},\omega) +
	\Ip\big(\alpha^\rH_k(\omega)\big) D^\rH_{kk}(\mathbf{r},\omega)
	\Big) \; .
\nonumber	  	 
\end{eqnarray}	
Here we assume that the principal-axis system of the magnetic polarizability 
tensor $\alpha^\rH(\omega)$ coincides with that of its electric counterpart. 
A particularly favorable case occurs for spheroidal nanoparticles 
({\em i.e.\/}, for rotational ellipsoids with two equal semi-axes $R_{\rm b}$ 
and a different third semi-axis $R_{\rm a}$), for which the polarizabilities 
$\alpha^\rE_k(\omega)$ and $\alpha^\rH_k(\omega)$ are known 
analytically~\cite{LaLiVol8}. In Fig.~\ref{fig:F_1} we plot the ratios 
$r_{\alpha}^{\rE/\rH} \equiv 
\Ip(\alpha_{\rm a}^{\rE/\rH}) / \Ip(\alpha_{\rm b}^{\rE/\rH})$ of 
the absorptivities along (a) and perpendicular to (b) the axis of rotation 
for spheroidal Au nanoparticles at the frequency $\oth = 10^{14}$~s$^{-1}$, 
roughly corresponding to the dominant thermal frequency at $T = 300$~K. 
We employ the Drude model 
\begin{equation}
	\varepsilon(\omega) = 
	1 - \frac{\omega_{\rm p}^2}{\omega^2 + \ri\gamma\omega}
\end{equation}
for the permittivity, with plasma frequency 
$\omega_{\rm p} = 1.4 \cdot 10^{16}$~s$^{-1}$ and relaxation rate 
$\gamma = 3.3 \cdot 10^{13}$~s$^{-1}$ for gold at room temperature, 
and keep the particle's volume constant when varying the aspect ratio 
$R_{\rm b}/R_{\rm a}$. As expected, one observes $r_{\alpha}^{\rE} \gg 1$ 
for prolate, rice grain-like particles ($R_{\rm b}/R_{\rm a} \ll 1$), 
whereas $r_{\alpha}^{\rH} \gg 1$ for strongly oblate, pancake-like particles 
($R_{\rm b}/R_{\rm a} \gg 1$), in which the magnetic fields can easily induce 
eddy currents. We expect that the model yields reliable results at least in 
the interval $1/5 \le R_{\rm b}/R_{\rm a} \le 5$ of the aspect 
ratio~\cite{HuthEtAl10}.

\section{Numerical model calculations}
\label{S_3}

\begin{figure}
\includegraphics[scale=0.30,angle=0]{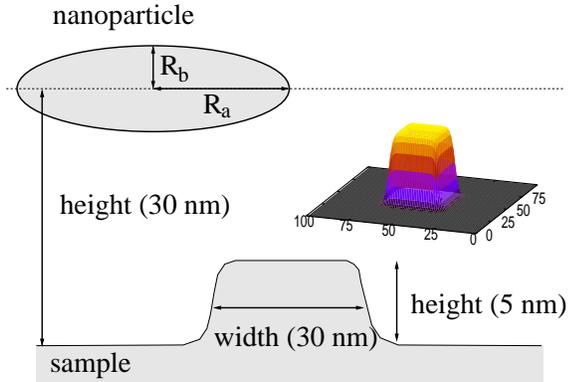}
\caption{Cross section, and 3D plot, of the surface structure employed in 
	the numerical model calculations. The orientation of the spheroidal 
	nanoparticle here corresponds to the scan of the near-field heat 
	transfer shown in the later Fig.~\ref{fig:F_5}.}			
\label{fig:F_2}	
\end{figure}

For our model calculations we take a substrate which is perfectly flat 
except for a square pad with 5~nm height, 30~nm width, and rounded edges, 
as depicted in Fig.~\ref{fig:F_2}. The coordinate system is oriented such 
that the $x$-$y$-plane coincides with the base substrate plane; the $x$- 
and $y$-axes being parallel to the pad's edges. We again assume that the 
permittivity of the substrate is described by the Drude model with parameters 
appropriate for gold.

\begin{figure}
\includegraphics[scale=0.35,angle=0]{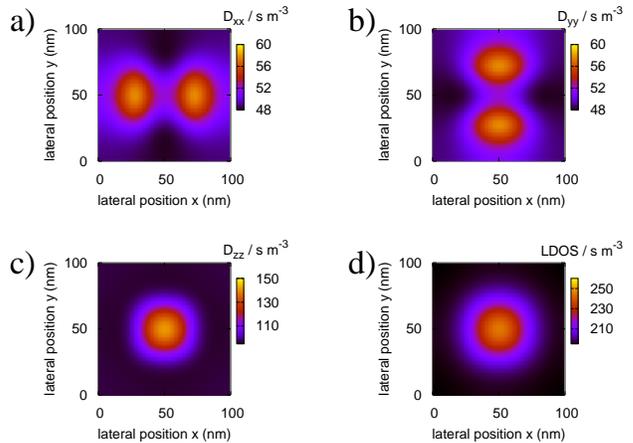}
\caption{a) -- c): ``Electric'' elements $D^\rE_{kk}(\mathbf{r},\omega)$
	($k = x,y,z$) for the model surface depicted in Fig.~\ref{fig:F_2}, 
	computed for $\oth = 10^{14}$~s$^{-1}$ at a distance of 30~nm 
	above the base $x$-$y$-plane. The dielectric properties of the 
	substrate are given by the Drude permittivity with parameters for 
	gold. d): Electric LDOS, as corresponding to the sum of a) -- c).} 
\label{fig:F_3}		
\end{figure}

\begin{figure}
\includegraphics[scale=0.35,angle=0]{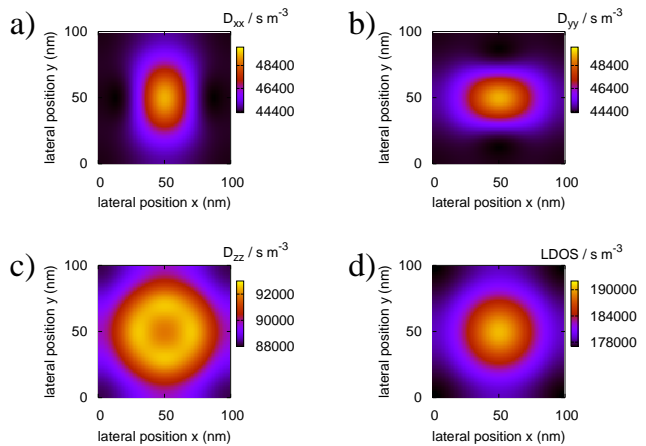}
\caption{As Fig.~\ref{fig:F_3}, for the ``magnetic'' elements 
	$D^\rH_{kk}(\mathbf{r},\omega)$. Observe that the scales 
	differ from those in Fig.~\ref{fig:F_3}.} 
\label{fig:F_4}		
\end{figure}

In Fig.~\ref{fig:F_3} a) -- c) we show the ``electric'' elements 
$D^\rE_{kk}(\mathbf{r},\omega)$ as defined by Eq.~(\ref{eq:DKK}), evaluated 
for $\omega = \oth$ at a distance of 30~nm above the base substrate plane. 
These elements have been computed by means of a perturbative approach to 
first order in the surface profile~\cite{BiehsEtAl08}. The sum of these 
three quantities gives the electric part of the LDOS, plotted in 
Fig.~\ref{fig:F_3}~d). Whereas the LDOS merely yields a blurred image of the 
underlying surface topography~\cite{KittelEtAl08}, the individual diagonal 
elements $D^\rE_{kk}(\mathbf{r},\omega)$ contain interesting directional 
information. The corresponding results for the ``magnetic'' elements 
$D^\rH_{kk}(\mathbf{r},\omega)$ are shown in Fig.~\ref{fig:F_4}.

For understanding the message contained in these figures, it is helpful to also 
consider the normalized eigenvectors $\mathbf{f}^{(j)}(\mathbf{r},\omega)$ of
the $3 \times 3$-matrix $D^\rE(\mathbf{r},\omega)$, and the corresponding
eigenvalues $d^{(j)}(\mathbf{r},\omega)$, together with their magnetic
counterparts~\cite{ZhangHenkel07}. Intuitively speaking, the eigenvectors
give the principal directions of the local field fluctuations; of course,
these directions are determined by the surface profile. The eigenvalues then
specify the strength of the partial fluctuation associated with the 
respective principal direction. In compact matrix notation, one now has
\begin{equation} 
	D^\rE(\mathbf{r},\omega) = S \, d \, S^{\dagger} \; ,
\end{equation}	
where the $j$th column of the matrix $S$ is given by the vector 
$\mathbf{f}^{(j)}(\mathbf{r},\omega)$, the matrix $S^{\dagger}$ is the adjoint 
of $S$, and $d$ is the diagonal matrix with $d^{(j)}(\mathbf{r},\omega)$ at 
the $j$th position on the diagonal line. Written out in components, this 
implies
\begin{equation}
	D^\rE_{kk}(\mathbf{r},\omega) = \sum_{j=1}^3 
	d^{(j)}(\mathbf{r},\omega) \, |f_k^{(j)}(\mathbf{r},\omega)|^2 \; .
\end{equation}	
The square $|f_k^{(j)}(\mathbf{r},\omega)|^2$ obviously specifies the weight 
with which the eigenvalue $d^{(j)}(\mathbf{r},\omega)$ contributes to the 
fluctuations along the $k$-direction. Thus, the individual diagonal elements 
$D^\rE_{kk}(\mathbf{r},\omega)$ express the distribution of the entire 
fluctuation strength, which in this language corresponds to the sum
$\sum_{j=1}^3 d^{(j)}(\mathbf{r},\omega)$ and thus is proportional to the 
electric energy density~(\ref{eq:EDE}), over the individual directions. 
Analogous remarks apply to the magnetic fluctuations. This kind of directional 
information is, in principle, contained in the thermal near field above a 
nano\-structured surface, and allows one to deduce properties of the structure 
itself. The experimental task is to actually extract such information from 
given laboratory samples, and the theoretical task is to devise methods for 
doing so.  

To this aim, we suggest a further development of the NSThM: The tip of this 
instrument is functionalized to act as a thermocouple, so that the small 
temperature change resulting from the energy deposited in or drawn out of 
the tip translates into a measurable voltage~\cite{MuellerHirschEtAl99,
KittelEtAl05,WischnathEtAl08}. If the active volume could be given the 
shape of a spheroid, the shape-dependence of the heat transfer expressed 
by Fig.~\ref{fig:F_1} could be exploited for assessing individual elements 
$D^{\rE/\rH}_{kk}(\mathbf{r},\omega)$ by means of Eq.~(\ref{eq:PSP}). 
Additional experimental handles would then be provided by the choice of
both the spheroid's orientation and its material. 
  
In order to demonstrate the feasibility of this concept, we compute the
near-field heat transfer between our model gold substrate with temperature
$\Ts = 300$~K and spheroidal nanoparticles cooled down to $\Tp = 100$~K. 
For metallic nanoparticles the magnetic contribution to the total heat 
transfer generally is much larger than the electric 
one~\cite{DedkovKyasov07,ChapuisEtAl08}, so that one has to select 
non-metallic sensors when aiming at the electric components 
$D^\rE_{kk}(\mathbf{r},\omega)$. Here we consider a rice grain-like 
particle with the permittivity of gallium nitride (GaN), described by 
the formula~\cite{BarkerIlegems73}
\begin{equation}
	\epsilon(\omega) = \epsilon_{\infty}
	+ \frac{S\omega_0^2}{\omega_0^2 - \omega^2 - \ri\Gamma\omega}
	- \frac{\omega_{\rm n}^2}{\omega^2 + \ri\gamma\omega}
\end{equation}
with parameters $\epsilon_{\infty} = 5.4$, $S = 5.1$, 
$\omega_0 = 1.0 \times 10^{14}\,\rm{s}^{-1}$,
$\omega_{\rm n} = 9.1 \times 10^{14}\,\rm{s}^{-1}$,
$\Gamma = 2.7 \times 10^{12}\,\rm{s}^{-1}$,
and $\gamma = 7.1 \times 10^{14}\,\rm{s}^{-1}$; 
the values of the inverse relaxation times $\Gamma$ and $\gamma$ have 
been properly adjusted to the particle's temperature. We choose the 
half-axes $R_{\rm a} = 40$~nm and $R_{\rm b} = 10$~nm, so that 
$r^\rE_\alpha \approx 23.4$ at $\oth$. These parameters clearly fall outside 
the regime where the dipole model~(\ref{eq:PSP}) can give quantitatively 
accurate results. We employ this model nonetheless, as computational tools 
for calculating the near-field heat transfer between nanoparticles and 
structured surfaces are not yet available, and stress the qualitative nature 
of the following results. 

\begin{figure}
\includegraphics[scale=0.30,angle=-90]{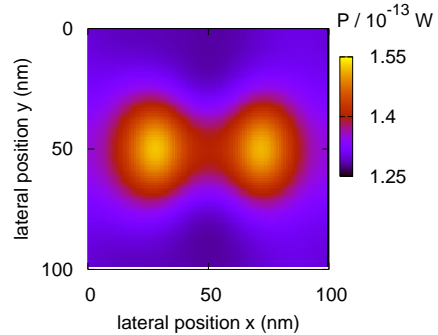}
\caption{Total heat transfer between a cooled spheroidal GaN nanoparticle
 	($\Tp = 100$~K) with $R_{\rm a} = 40$~nm, $R_{\rm b} = 10$~nm, and 
	axis of rotation parallel to the $x$-axis, and the gold structure 
	sketched in Fig.~\ref{fig:F_2} ($\Ts = 300$~K), assuming that the 
	particle is scanning the structure with a constant height of 30~nm.} 
\label{fig:F_5}		
\end{figure}

Figure~\ref{fig:F_5} shows the total heat transfer according to the
model~(\ref{eq:PSP}), with the particle oriented parallel to the $x$-axis 
at a constant height of $30$~nm above the base plane while scanning the 
substrate, as previously sketched in Fig.~\ref{fig:F_2}. In this case the 
heat flux mainly picks up the contribution proportional to 
$D^\rE_{xx}(\mathbf{r},\oth)$. As a consequence, the signal obtained here
closely resembles the previous Fig.~\ref{fig:F_3}~a). The slight differences 
between the two figures visible above the pad's center are explained by the 
fact that $D^\rE_{zz}(\mathbf{r},\oth)$ contributes significantly here.

\begin{figure}
\includegraphics[scale=0.30,angle=-90]{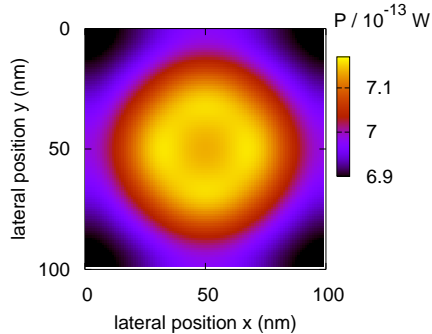}
\caption{Total heat transfer between a cooled spheroidal Au nanoparticle
 	($\Tp = 100$~K) with $R_{\rm a} = 10$~nm, $R_{\rm b} = 40$~nm, and 
	axis of rotation parallel to the $z$-axis, and the gold structure 
	sketched in Fig.~\ref{fig:F_2} ($\Ts = 300$~K), assuming that the 
	particle is scanning the structure with a constant height of 30~nm.}
\label{fig:F_6}		
\end{figure}

For assessing the magnetic contributions, we select a pancake-like
Au nanospheroid with $R_{\rm a} = 10$~nm and $R_{\rm b} = 40$~nm, giving 
$r^\rH_\alpha \approx 8.5$, and orient it along the $z$-axis. The total heat 
transfer then is dominated by $D^\rH_{zz}(\mathbf{r},\oth)$, as witnessed by 
the good agreement of Fig.~\ref{fig:F_6} with \ref{fig:F_4}~c). Thus,
here the recorded heat flux provides information on a component of the
magnetic correlation tensor.

\section{Conclusion}

In this paper we have outlined that individual diagonal components of the 
correlation tensors of fluctuating thermal fields close to the surface
of a given material can be measured by near-field thermal imaging, if the 
sensor employed for recording the heat flux is appropriately shaped. 
Evidently this proposal still is somewhat sketchy and idealized: On the 
theoretical side, there is an urgent need for accurate computational tools
for quantifying the near-field heat flux between arbitrarily shaped
nanoparticles and structured surfaces; from the experimental viewpoint,
the development of the required sensors poses a serious challenge. Thus,
there  may still be a long way from our present matter-of-principle 
considerations to laboratory applications.

Yet, there are at least two reasons for exploring this option. First, 
fluctuational electrodynamics is a macroscopic theory; it would be important 
to test its predictions even for probe-sample distances of a few nanometers 
and below, where its validity can no longer be taken for granted. Possible 
deviations form the macroscopic theory would then yield important novel 
insight into fundamental processes in dielectrics. For this program sensitive 
observables are required; we suggest that measuring individual diagonal 
components of  the near fields' correlation tensors, which give directional
weights to the total fluctuation strength, might yield more information than 
their traces, {\em i.e.\/}, the local density of states. Second, leaving 
theoretical considerations aside, the practical exploitation of the principles 
outlined in this note for nanoscale thermal imaging~\cite{KittelEtAl08} would 
provide novel tools for the diagnosis and characterization of nanostructures.

\begin{acknowledgments}
This work was supported by the Deutsche Forschungsgemeinschaft through 
Grant No.\ KI 438/8-1. Computer power was obtained from the GOLEM~I cluster
of the Universit\"at Oldenburg. S.-A.~B.\ gratefully acknowledges support  
from the Deutsche Akademie der Naturforscher Leopoldina under Grant No.\
LPDS 2009-7.
\end{acknowledgments}

\end{document}